\documentstyle[aasms4]{article}
\def\deg{\ifmmode^\circ _\cdot\else$^\circ _ \cdot$\fi }    %Degree sign%
\def\degg{\ifmmode^\circ \else$^\circ $\fi } 
\def\arcs{\ifmmode {'' }\else $'' $\fi}     %Arc seconds%
\def\arcm{\ifmmode {' }\else $' $\fi}     %Arc minutes%
\def\buildrel#1\over#2{\mathrel{\mathop{\null#2}\limits^{#1}}}
\def\mper{\ifmmode \buildrel m\over . \else $\buildrel m\over .$\fi}
\def\hper{\ifmmode \rlap.^{h}\else $\rlap{.}^h$\fi}
\def\sper{\ifmmode \rlap.^{s}\else $\rlap{.}^s$\fi}
\def\arcsper{\ifmmode \rlap.{' }\else $\rlap{.}' $\fi}
\def\arcmper{\ifmmode \rlap.{'' }\else $\rlap{.}'' $\fi}
\begin{document}
\def\fe{$\langle{\rm Fe}\rangle$~}
\title{The stellar and gaseous kinematics in NGC~253} 
\author{F. Prada\altaffilmark{1}, C.M. Guti\'errez\altaffilmark{1}, C. D. McKeith\altaffilmark{2}\altaffilmark{,3}}

\altaffiltext{1}{Instituto de Astrof\'\i sica de Canarias, 38200 La Laguna, Tenerife, SPAIN, (http://www.iac.es). E-mail: fprada@iac.es, cgc@iac.es}
\altaffiltext{2}{Department of Pure and Applied Physics, Queen's University of
Belfast, Belfast BT7~1NN, UK}
\altaffiltext{3}{Deceased}
%\lefthead{Prada, Guti\'errez \& McKeith}
%\righthead{The stellar and gaseous kinematics in NGC~253}

\begin{abstract}

This paper presents  observations with intermediate spectral 
and spatial resolution along the major and minor
axes of the starburst galaxy NGC~253. The spectral ranges analyzed are
in the region of the stellar MgI$b$ ($\sim 5175$ \AA), the near IR
CaII triplet ($\sim$ 8550 \AA) absorption features, and 
the region of the H$\alpha$  emission line.
We have compared the shape of the stellar features with those of
reference stars and determined the line-of-sight velocity distribution (LOSVD)
of the stellar component by using a two-dimensional Gaussian
decomposition algorithm, and show for the first time the rotation curve of the
stellar component in NGC~253. Comparing the recesion velocity curves of the
gas and stars, it is seen that the stellar component has a decoupled
kinematics respect to the gas, displaying a shollower velocity gradient
and larger velocity
dispersion that the gas in the inner regions. The minor axis
kinematics together with the kinematics across the central 40$''$ along
the major axis, suggest the presence of a rotating body with a 
kinematically misaligned
axis respect to the main disk of the galaxy. The asymmetries in the LOSVD
along the minor axis together with the steep velocity gradient of the gaseous
component suggest a merger scenario to explain the these
kinematically signatures. The enclosed mass in the
central regions is computed to be (2.4$\pm 0.5)\times
10^7$ M$_\odot$ for a radius of $r=0.7''$ (10 pc). A double gaseous
component in the central 6 arcsecs is detected from the 
[SIII] $\lambda$ 9069 \AA\ data along the minor axis; this seems
to be the signature of a superbubble due to a supernova rate of 0.05 yr$^{-1}$.

\end{abstract}


%\newpage

\section{INTRODUCTION}

The well known starburst galaxy NGC~253 (e.g. Rieke et al. 1980),
classified as type SAB(s)c (de Vaucouleurs et al. 1991) has been
subject of many detailed kinematical studies of its gaseous component.
These studies include long-slit optical spectroscopy (Ulrich 1978;
Mu\~noz-Tu\~non,Vilchez \& Casta\~neda 1993; 
Arnaboldi et al. 1995,hereinafter ARN95),
which shows large deviations from non-circular motions. These deviations
have been attributed to the outflow of gas due the presence of massive
star formation (e.g. Heckman et al. 1990), as well as the presence of
heavy obscuration in the central regions (Scoville et al. 1985; Sams et
al. 1994). The large amount of extinction seems to be the reason for
the steeper rotation curve obtained with  near-IR emission lines compared 
to  optical lines
(Puxley \& Brand 1995; Prada et al. 1996). The infrared measurements are 
in agreement
with the $^{12}$CO (J=1-0) velocity field derived by Canzian et al.
(1988). The most complete analysis of the gaseous component
in the innermost regions has been done using the radio recombination
H92$\alpha$ line by Anantharamaiah \& Goss (1996) (hereinafter AG96).
They determined a velocity field in the central 150 pc displaying a
distinct gaseous kinematics subsystem, exhibing rotation in a plane
perpendicular to the galactic disk, and an inner region with possible
counterrotation in the plane of the disk. The authors suggest that this
result may be indicative of a secondary bar inside the known primary
bar, or alternatively, a signature of a merger or an accretion event
during the history of the galaxy.

The analysis and comparison of the stellar kinematics with the
gaseous motions would be of great interest in determining the nature of this
peculiar kinematics found in the central regions of the galaxy.
However, up to now the information on the stellar kinematics of NGC~253
is poor and only the velocity dispersion in the nucleus is known (Lutz
\& Prada 1995; Oliva et al. 1995). The study presented in this paper
reports the results of spectroscopic observations taken in the region
of the MgI$b$, the CaII IR triplet stellar features, and the H$\alpha$
emission line along the major
and minor axes of NGC~253. 

The paper is structured as follows: Section
2 presents the observations and data reduction, Section 3 analyzes
the gaseous kinematics and the starbust in the center of the galaxy, and
 Section 4 determines the stellar kinematics and evaluates the 
 enclosed mass in the nuclear region. Section 5 is  dedicated to the
comparison between the stellar and gaseous kinematics.

\section{OBSERVATIONS AND DATA REDUCTION} 

During several campaigns, we obtained long-slit spectra  along the major
(p.a. 52$^{\circ}$) and minor (p.a.  321$^{\circ}$) axes of NGC~253
with the double arm ISIS spectrograph on the 4.2 m William Herschel
Telescope at the Observatorio del Roque de los Muchachos, La Palma,
Spain. We used a 1124$\times 1124$ Tek CCD with 0.36$^{\prime
\prime}$ pixel$^{-1}$ in the red and blue arm of ISIS
respectively.  The spectral dispersion was 0.78 \AA~per pixel (with a
FWHM=2 pixels); CuAr and CuNe  lamps were used for spectral
calibration.  The observations are summarized in Table~1. In the
observations of November 1995, the slit was $0.9''$ wide, slightly
undersampling the $\sim 0.9''$ seeing, and was placed along the major
axis centered on the apparent optical nucleus. In the observation of
July and August 1996 the slit was placed at the dynamical center given
by AG96 with a slit width of 1.3$^{\prime \prime}$ and seeing of $\sim
1.2^{\prime \prime}$. The slit was 4$^{\prime}$ long in all the runs.
In each of these runs, standard stars with spectral types
ranging from G5III to K4III were also observed. These stars were used
as templates to determine the stellar line of sight velocity
distribution (LOSVD) of NGC~253 as it will be explained in next sections.
Standard data reduction including bias subtraction, flat field
correction and wavelength calibration was carried out using the
IRAF$\footnote{IRAF is distributed by National Optical Astronomical
Observatories, which is operated by the Association of Universities for
Research in Astronomy, Inc., under contract with the National Science
Foundation}$ package. The sky lines were subtracted using template
spectra from the ends of the slit. A distance of 3.4 Mpc (Sandage \&
Tammann 1975) for NGC~253 was adopted, yielding a scale of 16.5 pc
arcsec$^{-1}$.

\section{THE GASEOUS KINEMATICS AND STAR FORMATION ACTIVITY}

Figure~1 shows the gaseous position-velocity curves along the major
axis of the galaxy, determined by fitting Gaussian profiles to the
[NII] $\lambda$ 6584 \AA~and the [SII] $\lambda \lambda$
6717 6731 \AA\, H$\alpha$ emission lines. The radial velocities are 
heliocentric and the zero position
for the spatial axis comes from the continuum peak at $\sim$ 6500 \AA. 
The systemic
velocity of $V_{\rm HEL}=$230 $\pm$ 10 km s$^{-1}$ is in agreement with
previous measurements (e.g. ARN95). The results for the 
three species 
are in
agreement and  show that the center of symmetry for the gas rotation 
curve is
shifted 11$''$ SW from the continuum peak as was first reported
by ARN95. Figure 2 compares the 
$[$SIII$]$ $\lambda$ 9069
\AA\ velocity curve (measured in November 1995) 
with the near-IR H$_2$ $\lambda$=2.1218 $\mu$m
position-velocity curve determined by Prada et al. (1996), and with
recent measurements  by AG96 using the radio recombination line
H92$\alpha$. There is a good agreement between the three species except
for the SW of the nucleus suggesting that this side is especially
dusty.  This velocity difference has been also
found by Prada et al. (1996) comparing near-IR data with optical data,
and was attributed to an optical depth effect due to dust extinction
(see Section 5).

Figure 3 compares the minor axis position-velocity curve of the [SIII]
emission line with the H$\alpha$ data by Ulrich (1978) and Heckman et
al. (1990), as well as for the measurements by AG96 using H92$\alpha$.
The [SIII] spectra show two components in the central 6$''$: the
brighter component has the systemic velocity (V$_{HEL}$=230 km s$^{-1}$)
 while the fainter one is
blueshifted 140 km s$^{-1}$.  Further out from the center, our data
agree with the Ulrich (1978) data, whilst the data by Heckman et al
(1990) given only for the SE, show two components beyond 10$''$ from
the center. We believe that the deeper penetration through the dust
of the near-IR
wavelengths allow us to see into the equator of the galaxy the two
components that Heckman et. al (1990) and Schulz \& Wegner (1992)
attribute to a starburst-driven superwind at the surface of a cone or
bubble. However, our [SIII] data do not agree with the very steep
H92$\alpha$ velocity gradient reported by AG96 in the inner 4$''$ (see
Figure 3); this will be discussed in detail in Section 5.

Figure 4 presents the velocity dispersion curve along the major axis of 
NGC~253 for the
optical and the [SIII] $\lambda$ 9069 \AA~emission lines. The optical 
velocity dispersion curve shows a depression
 with a minimum value of 65 km s$^{-1}$ at the dynamical center
 (11$^{\prime \prime}$ from the continuum peak), and reaches $\sim 100$
 km s$^{-1}$ at 10$''$ at either side of this. For the [SIII],
 the velocity dispersion reaches a
value of 85  km s$^{-1}$ at 3.5 SW from the continuum peak 
and decreases slowly towards larger 
radii. The value
of the velocity dispersion at the dynamical center is in agreement with
the measurements by Oliva et al. (1996) using Br$\gamma$ for a central
aperture of 4$''$x4$''$.

The two components present in the [SIII] emission lines in the central
4$''$ along the minor axis of NGC~253, can be interpreted as a
large-scale expansion or superbubble due to supernova (SN) events. The
measurement of the gas expansion velocity at the position where the
slit interset the disk of the galaxy, namely the expansion velocity of
the superbubble, can be used to constrain the SN rate in the nuclear
regions. Colina \& P\'erez-Olea (1992) consider a SN rate in the nuclear 
region of
0.05 yr$^{-1}$, in  agreement with
the value  obtained by
Heckman et al. (1990). Taking
our measurements, the expansion velocity is $\sim$ 75 km s$^{-1}$
deprojected using an inclination of 78.5$^{\circ}$ (Pence 1981) and
 the density of the medium in the bubble, n$_o$=630 cm$^{-3}$ (Heckman
et al. 1990), we estimate a radius of 130 pc for the superbubble in the
disk of the galaxy. The radius can be estimated 
from the equations given by McLow \& McCray (1988) for the mechanical
luminosity of a superbubble  $L_{SN}\sim
3.6\times 10^{29}n_o$ (cm$^{-3}$)$R^2({pc})V^3$({km s$^{-1})^3$. As
 the superbubble should be located
interior to  the edge-on nuclear ring of 20$''$ (330 pc) in diameter
reported by ARN95, this introduces a constrain in
in the SN rate with values much smaller than the one derived by 
 by Rieke et al. (1988) and Forbes et al.  (1993) who estimates 
a SN rate of 0.1 or 0.2 yr$^{-1}$. For this reason we believe that the average 
SN rate in the central 300 pc is $\sim$0.05 yr$^{-1}$. The value of 
0.1-0.2 yr$^{-1}$
could be more local since it was determined from the presence of few
radio sources (Rieke et al. 1988), and the [FeII] luminosity in the
central 50 pc (Forbes et al. 1993) of the galaxy. Finally, we would like to 
mention
that all the authors agree for a SN rate of $\sim$0.3 yr$^{-1}$ in the
nucleus of M82.

\section{THE STELLAR KINEMATICS AND ENCLOSED MASS}

We have determined the LOSVD along the major and minor axis of NGC~253
from the long-slit MgI$b$ (5175 \AA) and CaII IR triplet (8550 \AA)
absorption spectra by means of a two-dimensional unresolved Gaussian
decomposition algorithm. This method has been applied succesfully to
determine the stellar kinematical structure of NGC 7331 (see Prada et al.
1996). Basically, this approach considers that the spectrum of the
galaxy in a given angular position is the result of a {\it mean}
stellar spectrum broadened by the internal kinematic structure of the
galaxy and by the observational-experimental configuration. The
intrinsic broadening represents the convolution of this mean stellar
spectrum with the LOSVD.  Additionally the instrumental configuration
(slit-width, etc), and the atmosphere (seeing) sliglthly
 broaden the spectrum in
the plane of the detector, introducing correlations between neighbouring
pixels along both the spectral and the spatial axes respectively. We
parametrize the LOSVD as a sum of two-dimensional Gaussians in the
plane {\it velocity-angular position}. The dispersion of the Gaussians
reflect the resolution in both axes, {\it i.e.} spectral and spatial
resolution respectively.  The dispersions used for the observations
presented here, were  $\sim $2 pixels and $\sim $1.5 pixels
 in the spectral and spatial axes respectively, with relative
separation 3 and 2 pixels. The algorithm makes a two-dimensional $\chi
^2$ minimization of the difference between the observed and the
modelled spectra using a Netwon-Raphson iterative scheme. With respect to
the previous algorithm by Kuijken \& Merrifield (1993), the
two-dimensional decomposition takes into account the correlation
between adjacent angular positions mentioned above. The use of this
additional information  produces little change in regions of moderate o
high signal to noise ratio (SNR)  but might be important in regions far
away of the galactic center.

The spectral regions analyzed were $5120-5290$ \AA\ and $8450-8700$
\AA\ along the major and minor axes. As velocity template for the
November 1995 observations, the K0 giant HR218 star was used. The K4III
HD 213947 and the G5III HD4388 stars were used for the July and August
1996 runs respectively. Figure 5 shows the spectra in several positions
along the major axis in the region of the stellar MgI$b$ feature. The
emission feature at $\sim 5200$ \AA~corresponds to the gaseous line [NI]
$\lambda \lambda $ 5198.5 5200.7 \AA. Also plotted are the modelled
spectra obtained by convolution of the derived LOSVD with the spectrum
of the velocity template (excluding the [NI] feature). Several
absorption features including the main components of the MgI$b$ are
reconstructed indicating the reability of the algorithm, and that the
stellar reference spectra is a good representation of the mean stellar
spectrum in the galaxy. Figures 6 and 7 show the equivalent plots in the
region around the CaII IR triplet along the major and minor axes
respectively. In the case of the major axis, the spectra correspond to
the August 1996 run. In both cases the CaII line at 8662 \AA\ has been
excluded from the fit because it appears strongly contaminated by
sky emission. The reconvolved spectra are very similar using template
stars of types from G5III to K4III.  Exhaustive analyses of the
performance of the algorithm, including Monte Carlo simulations will be
presented in a forthcoming paper (Guti\'errez \& Prada 1997).

The LOSVD along the major axis obtained from the MgI$b$ and CaII IR
triplet are symmetrical and show a single component. Figure~8 shows the
radial velocities along the major axis; these have  been obtained by
fitting a single Gaussian to the LOSVD. These radial velocities are
heliocentric and the zero position have been chosen from the 
continuum peak in each case. The MgI$b$ velocity curve goes from
$-90''$ to $+120''$ from the continuum peak. Overplotted is a linear
fit which reproduces adequately the general trend of the curve and
which has a velocity gradient of 1.7 km s$^{-1}$ arcsec$^{-1}$.
Separation of this trend seems to occur in the region from $\sim +40''$
to $+90''$. In the case of the CaII triplet, the data of August 1996
allow to follow the velocity curve in a large spatial range (from
$-25''$ to $+25''$ from the continuum peak). Also shown is a fit
to the CaII data of a
circular velocity motion (see Bertola {\it et al.} 1991). The CaII
velocity curve shows a steeper gradient compared to the MgI$b$ curve
for the SW side probably due to the
presence of dust in the SW side of the nucleus (Scoville et al. 1985;
Sams et al. 1994).

Figure~9 presents the CaII IR triplet velocity curve for the minor axis
obtained by fitting a single Gaussian to the LOSVD. 
The curve
shows a  velocity gradient which is not expected from circular motions
along the main plane of the galaxy. Furthermore, the LOSVD shows some
degree of asymmetry towards the blueshift in the inner 10$^{\prime
\prime}$, as it is shown in Figure 10.

Figure~11 shows the stellar velocity dispersion curve from the CaII IR
data along both axes. The velocity dispersion in the major axis has a
value of 90$\pm$5 km s$^{-1}$ at $5''$ SW from the continuum peak which
is in agreement with the values that Oliva et al.  (1996), and Lutz \&
Prada (1995) give using the CO 2.3 $\mu$m absorption feature for a
central aperture of 4$''$$\times$4$''$. Beyond $\sim$ 10$''$ the 
velocity dispersion decreases down to 70 km s$^{-1}$. The velocity dispersion
along the minor axis reaches a value of 110$\pm$20 km s$^{-1}$ in the
central 10$''$. This can be attributed to the presence of two
kinematically distinct components in the inner regions as revealed by the
LOSVD.

Assuming that the velocity dispersion accounts for all the
gravitational support (see McGinn et al. 1989)  and following the
notation of Gaffney et al. (1993) the enclosed mass for a given radius
is \begin{equation} M(r\le r_o)=\frac{r\,\sigma^2(r)(A+B+C)}{G}.
\end{equation} where $A=-d\ln n(r)/d\ln (r)$ , $B=-d\ln \sigma ^2
(r)/d\ln r $, and $C=V^2(r)/\sigma ^2 (r)$. To determine $A$ we have
adopted a surface brightness profile $\mu (r)\propto r^{-0.8\pm0.2}$,
valid for 1$''\le r\le 5''$ determined from near-IR observations (Sams
et al. 1994). Hence, we can obtain the flux density $F(r)\propto
r^{1.5}$ and therefore  $n(r)\propto r^{-1.5}$. $B$ and $C$ were
determined from the results shown in Figures 8 and 11. Figure~13 shows
the enclosed mass at different radii: for $r=0.7''$ (10 pc),
we obtained an enclosed mass of (2.4$\pm 0.5)\times 10^7$ M$_\odot$.
This value is in agreement with the enclosed mass found by Gaffney et
al. (1993) for M82: $M (r\le 7.5$ pc)=$(3\pm1)\times 10^7$ M$\odot$.

\section{THE DICHOTOMY OF THE STELLAR AND GASEOUS KINEMATICS}

The direct comparison of our measurements of the optical
 stellar MgI$b$ and gaseous (e.g. H$\alpha$) position-velocity curves
 reveals a clear
dichotomy between the kinematics of the gas and the stars in the
central 650 pc of NGC~253 (see Figure~12). The effects of dust
extinction in determining the line-of-sight velocities are very small
for the two spectral lines mentioned above to explain such velocity
differences between gas and stars (see Prada 1995).  However, in the SW side of the nucleus, the heavy obscuration
(A$_V$$\sim$30 mag) reveals a steeper velocity gradient for the stellar
CaII near-IR feature, which is shallower than the H$\alpha$ velocity
gradient, the [SIII] line, the near-IR emission lines (Puxley \& Brand
1995; Prada el at. 1996), and the radio emission lines (AG96) in this
region, but steeper than the MgI$b$ velocity curve.

Adopting the larger wavelength velocity data, i.e. the CaII IR triplet
absorption feature and the emission H92$\alpha$/H$_2$ $\lambda 2.13$
$\mu$m lines as the intrinsic (i.e. extinction free) stellar and
gaseous kinematics of NGC~253, we conclude that the motions of the
stars in the central regions are not coupled with the gas. We think
that the velocity difference could be a consequence of the distribution
of the gas in a ring. This seems to be confirmed by the high resolution
measurements made  by ARN95 who claim the existence
of an edge-on nuclear ring due to the presence of an Inner Linblad
Resonance at 300 pc from the center. They believe that the bar is the
responsible for the gas fueling of the nuclear starburst, and therefore
will influence the dynamics of the gas in the central regions. This
suggests that the gas is distributed in a ring with ongoing star formation
and that the stars will trace the kinematics of the regions inside the ring.
This is reinforced by the larger value of the stellar velocity
dispersion with respect to the gas: 90 and 65 km s$^{-1}$ respectively. This
difference also occurs is other galaxies as reported by Terlevich et
al.  (1990) from observations of the CaII IR triplet. They  find that
the only objects which have emission lines substantially narrower than
absorption lines are starburst galaxies. Similar values than for NGC 253, 
have been found in the nucleus of M82 (Gaffney et al.
1993); in this galaxy, also the slope of the stellar velocity curve for
the central regions is shallower than  the ionized gas as reported by
Gaffney et al. (1995). These authors conclude that the stellar peak in
M82 may be an older bulge population (Gaffney \& Lester 1992; Gaffney
et al. 1993), which might also be the case of NGC~253 given the
remarkably similarity with the dynamics of M82.

The kinematics along the minor axis of NGC~253 reveals the presence of
an ordered motion which appears well described by a circular motion
with $\Delta$V$_{max}$$\sim$40 km s$^{-1}$
(see Figure~9). The minor axis kinematics together with the kinematics
across the central 40$''$ along the major axis which has 
$\Delta$V$_{max}$$\sim$70 km s$^{-1}$ suggest the presence of
a central rotating body with a misaligned axis respec to the main rotation
axis of the galaxy. According to the shape of the LOSVD, this body might
contains a decoupled kinematical component in its center probably due
to the accretion of material, either in the form of stars or gas which
tiggered the starburst.

The counterrotation reported by AG96 in the inner 2$''$
 along the major axis is also present in our
[SIII] data, although there is a velocity discrepancy in the SE regions
due to dust extinction as discussed in Section 3. The fact that our
[SIII] data match the H92$\alpha$ velocity data along the major axis,
suggests that dust extinction is not very high in the inner 4$''$ (see
Figure~2). Therefore, we do not understand, if this is the case, the
very step velocity gradient found by AG96 along the minor axis.
Furthermore our [SIII] minor axis data show two velocity components.
We believe that the discrepancy between the two data sets can be
interpreted as follows: by fitting a single Gaussian to the [SIII]
emission lines we can match the H92$\alpha$ data by shifting
3.5$^{\prime \prime}$ NE our data. This would suggest that the
H92$\alpha$ velocity gradient reported by AG96 might represent an
artefact due to the outflow of gas. On the other hand, it could well
happen that we placed the slit not exactly at the dynamical center,
since our positioning error was of 1.25$''$, and the rapid rotation
seen in the H92$\alpha$ data is a kinematic subsystem that might have
the same origin than the structure seen in the stellar LOSVD along the
minor axis.

\acknowledgements This research is based on observations obtained at
the WHT, operated by the ING at the Observatorio del Roque de los
Muchachos at La Palma of the IAC. We thank E. Perez and  R. F. Peletier
for helpful discussions. The spectroscopic  observations were performed
on service nights in November 1995 and July and August 1996.

This paper is dedicated to the memory of C. D. McKeith, who contributed
much to this work before his untimely death on Summer 1996.

\newpage

\subsection*{FIGURE CAPTIONS}

\figcaption[Prada.fig1.eps]{The optical gaseous position-velocity curves
along the major axis of NGC~253, for the H$\alpha$ (bottom), [NII]
$\lambda$ 6584 \AA~ (middle), and the [SII] $\lambda \lambda$ 6717,
6731 \AA~(top) emission lines. In the middle pannel the 
relative intensity of the continuum around 6500 \AA~along the major
axis is also plotted.}

\figcaption[Prada.fig2.eps]{A comparison among the position-velocity
curves along the major axis of NGC~253 of the [SIII] $\lambda$ 9069
\AA, the near-IR H$_2$ $\lambda$=2.1218 $\mu$m 
determined by Prada et al. (1996), and the measurements by
Anantharamaiah \& Goss (1996) using the radiorecombination line
H92$\alpha$. The dashed lines represents the relative intensity of the
continuum around 8750 \AA.}

\figcaption[Prada.fig3.eps]{The  position-velocity curve along the minor
axis of NGC~253 for our [SIII] $\lambda$ 9069 \AA\ data,the H$\alpha$
data by Ulrich (U78) and Heckman et al. (HAX90), and the  H92$\alpha$
by Anantharamaiah \& Goss (1996). The central region is plotted
magnified on the right. When present, the filled symbols refer to the
fainter component. The circles in the magnified plot  represent a
single Gaussian fit to the [SIII] emission lines.}

\figcaption[Prada.fig4.eps]{The velocity dispersion curve along the
major axis of NGC~253 for the H$\alpha$ (top) and the [SIII]
 $\lambda$ 9069 \AA\ (bottom)
emission lines.}

\figcaption[Prada.fig5.eps]{The spectra of NGC~253 along the major axis
in several positions in a region centred at 5200 \AA. The main
absorption line is the MgI$b$ features. The emision at $\sim 5200$
\AA~corresponds to the [NI] doublet gaseous line. The heavy lines show
the reconvolved spectra obtained using our algorithm (see the text).}

\figcaption[Prada.fig6.eps]{The same than in the previous plot but in
the region of the CaII IR absorption triplet ($\sim 8500$ \AA). Only
the two CaII features  at shorter wavelengths (heavy lines) were used
to determine the LOSVD.}

\figcaption[Prada.fig7.eps]{The same than in the previous plot but the
CaII velocities along the minor axis.}

\figcaption[Prada.fig8.eps]{Heliocentric radial velocities along the
major axis of NGC~253 obtained from the MgI$b$ (bottom), and the CaII IR triplet (top) absorption features
using the method described in the text. The solid lines
represent a circular motion fit to the caII data and a linear fit to the MgI
 data respectively. The dashed lines represent the relative intensity of the
continuum at $\sim 5000$ and  $8750$ \AA respectively.}

\figcaption[Prada.fig9.eps]{The same than the previous figure but for the 
CaII data along the minor axis.}

\figcaption[Prada.fig10.eps]{The stellar LOSVD along the minor axis in the 
central regions of NGC~253. Note a certain degree of assymetry towards the 
blueshift.}

\figcaption[Prada.fig11.eps]{The stellar velocity dispersion along the major 
and minor axis of NGC~253 (top and bottom respectively). The dispersions were 
obtained according the LOSVD determined from the CaII absorption features 
(August 1996 run).}

\figcaption[Prada.fig12.eps]{A comparison between the velocity-position curves 
for the gas (thin line) and stars (thick line) along the major axis of NGC~253 
obtained from the analysis of the H$\alpha$ and CaII lines respectively. The 
dashed line is a linear fit to the MgI$b$ stellar curve. The zero position in
the spatial axis corresponds to the dynamical center.}

\figcaption[Prada.fig13.eps]{The enclosed mass in NGC~253 as a function of the 
radius. The points and the error bars represent an estimate assuming that the 
velocity dispersion accounts for all the gravitational support. The solid line
is a linear fits.}

\newpage

{\tiny
\begin{table}
\caption{Observations}
\medskip
\begin{tabular}{lcccccc}
\hline
\noalign{\vspace {0.1cm}}
Date & Exposure time (s) & P.A. (\degg) & Spectral range (\AA) &Main features & Template stars (Spectral Type) \\
\noalign{\vspace {0.1cm}}
\hline
\noalign{\vspace {0.1cm}}
 November 95 & 1800 & 52 & 8300-9100  &CaII, [SIII] $\lambda$ 9069 \AA & HR 218 (K0 III)  \\
November 95 & 1800 & 52 &4800-5600  & MgI$b$, H$\beta$, [0III] $\lambda$ 5007 \AA & HR 218 (K0 III)\\
November 95 & 1800 & 52 & 6100-6900  & H$\alpha$, [NII] $\lambda$ 6584 \AA, [SII] $\lambda$ 6717 6731 \AA & HR 218 (K0 III)\\
July 96 & 1800 & 321 &8300-9100 & CaII, [SIII] $\lambda$ 9069 \AA & HD 213947 (K4 III) \\
August 96 & 1800 &  52 & 8300-9100 & CaII, [SIII] $\lambda$ 9069 \AA  & HD 4388 (G5III) \\
\noalign{\vspace {0.1cm}}
\hline
\end{tabular}
\end{table}
}

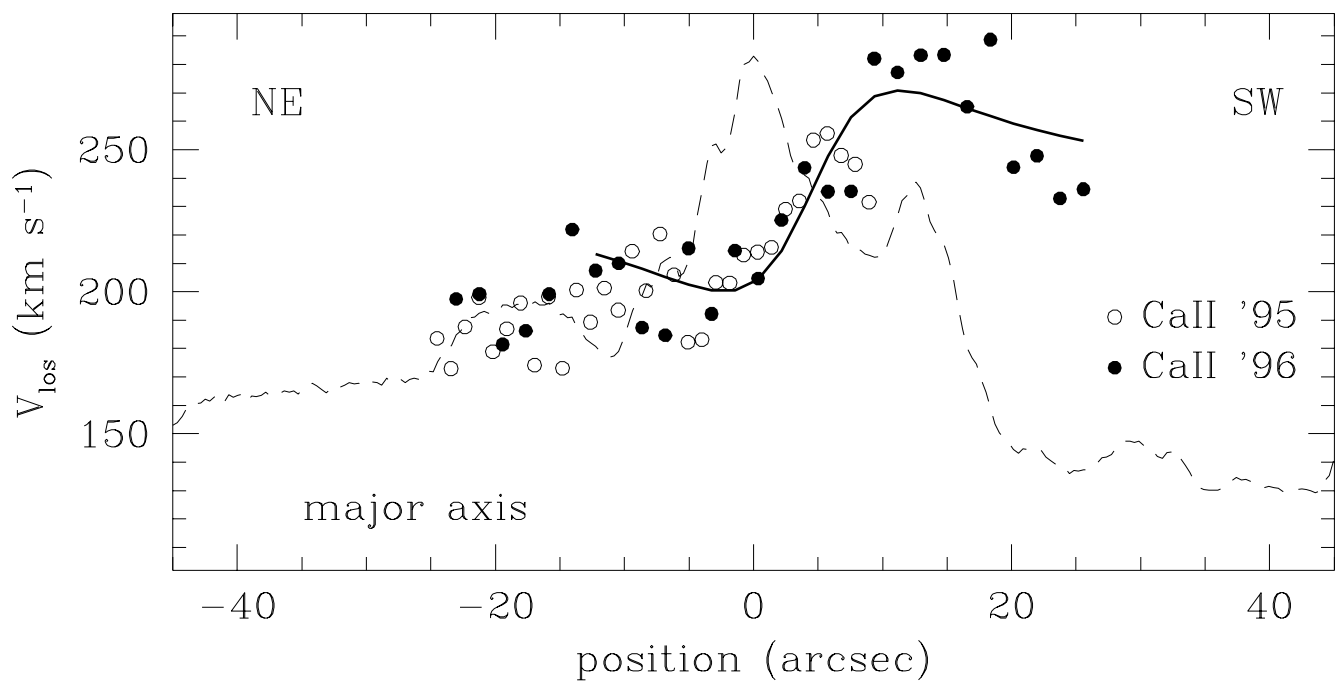
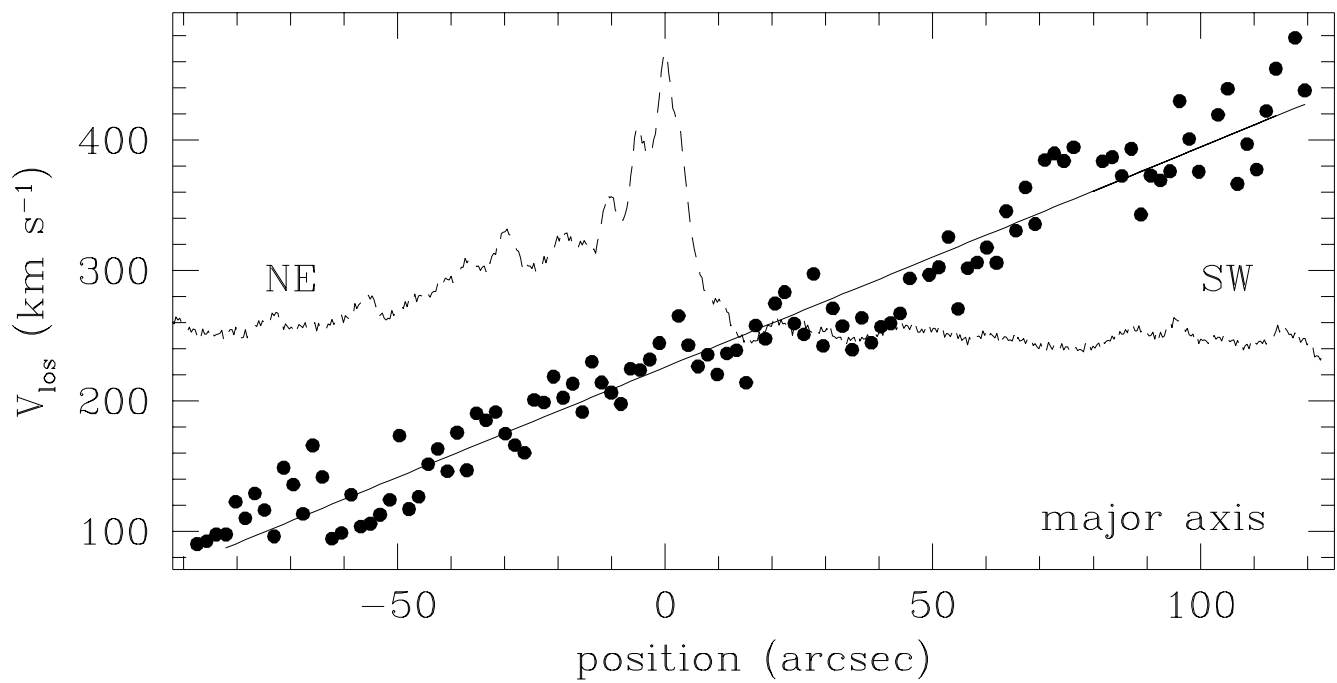
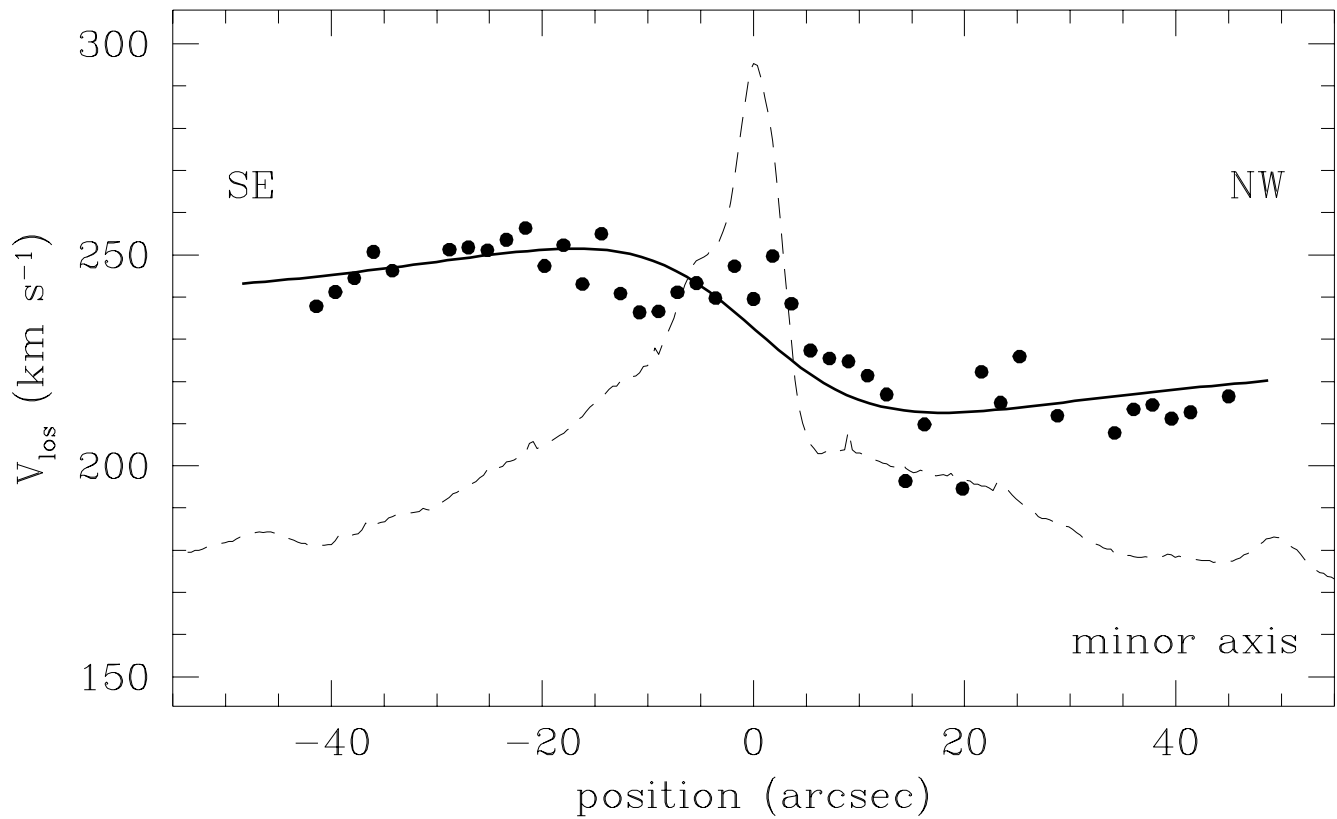
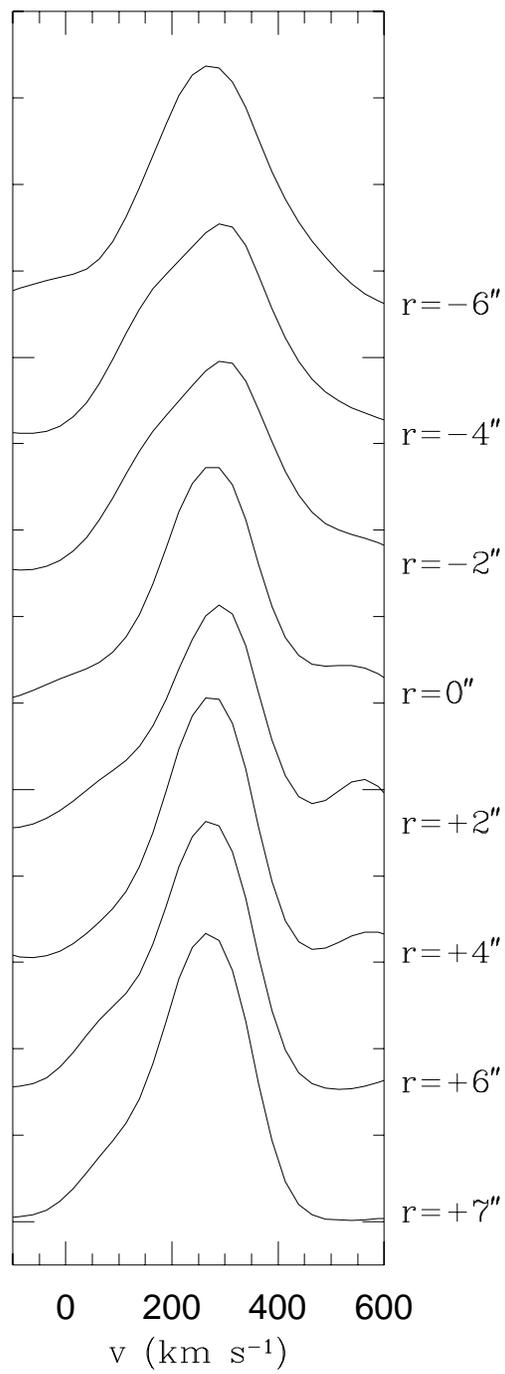
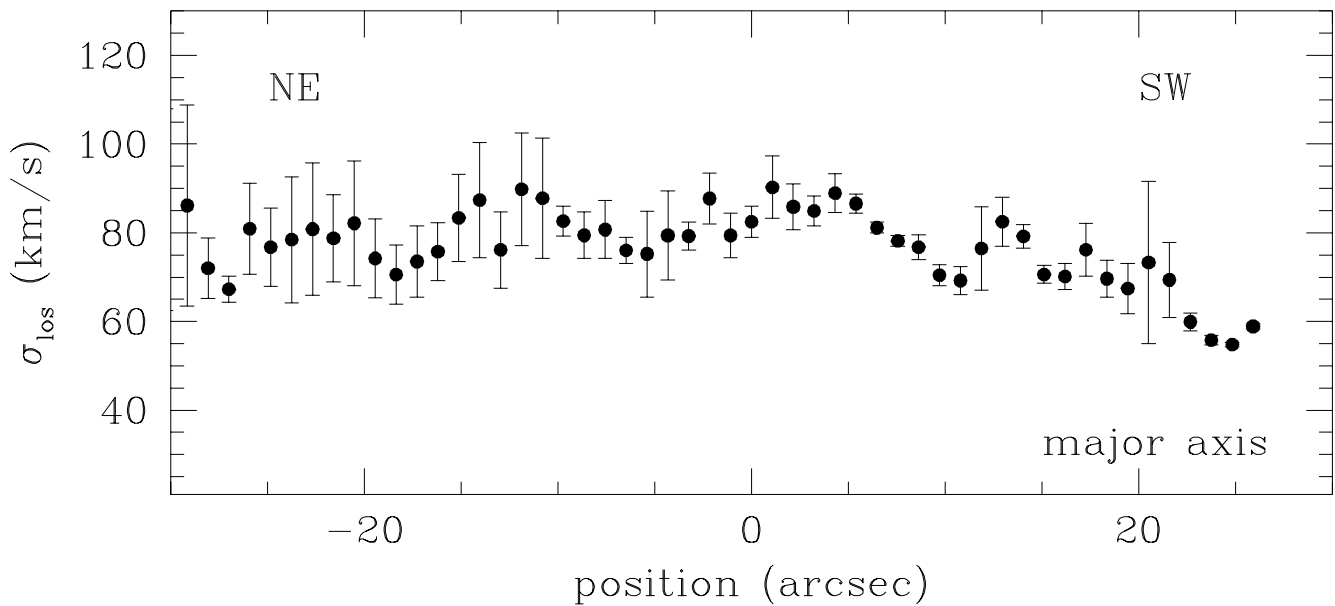
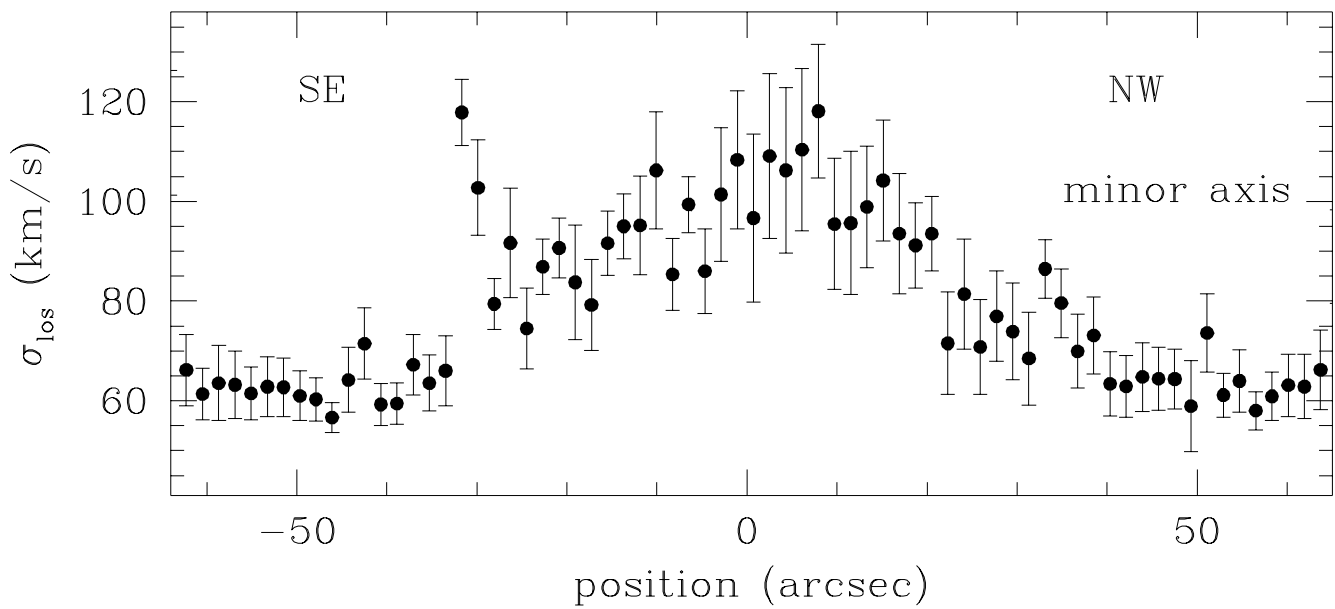
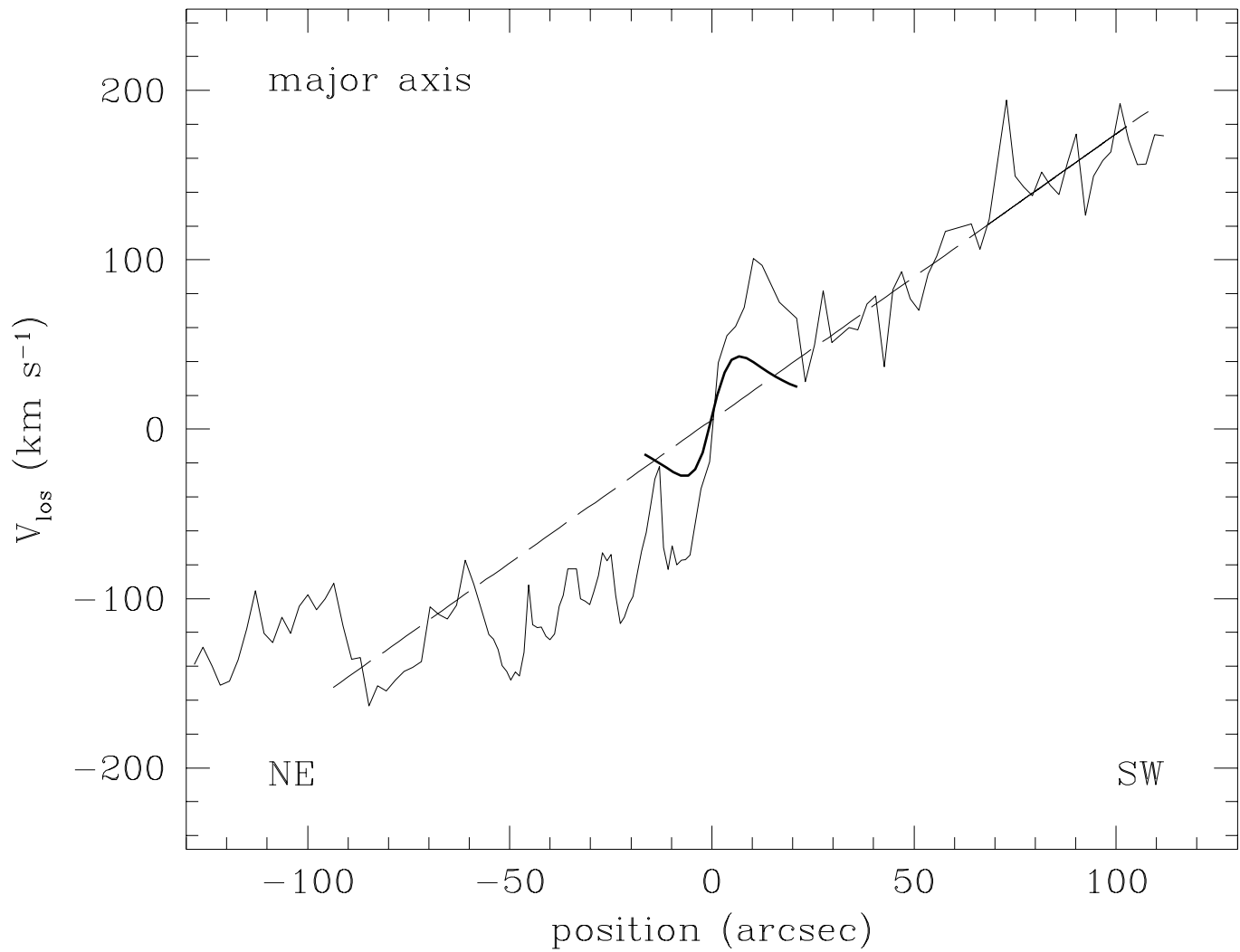
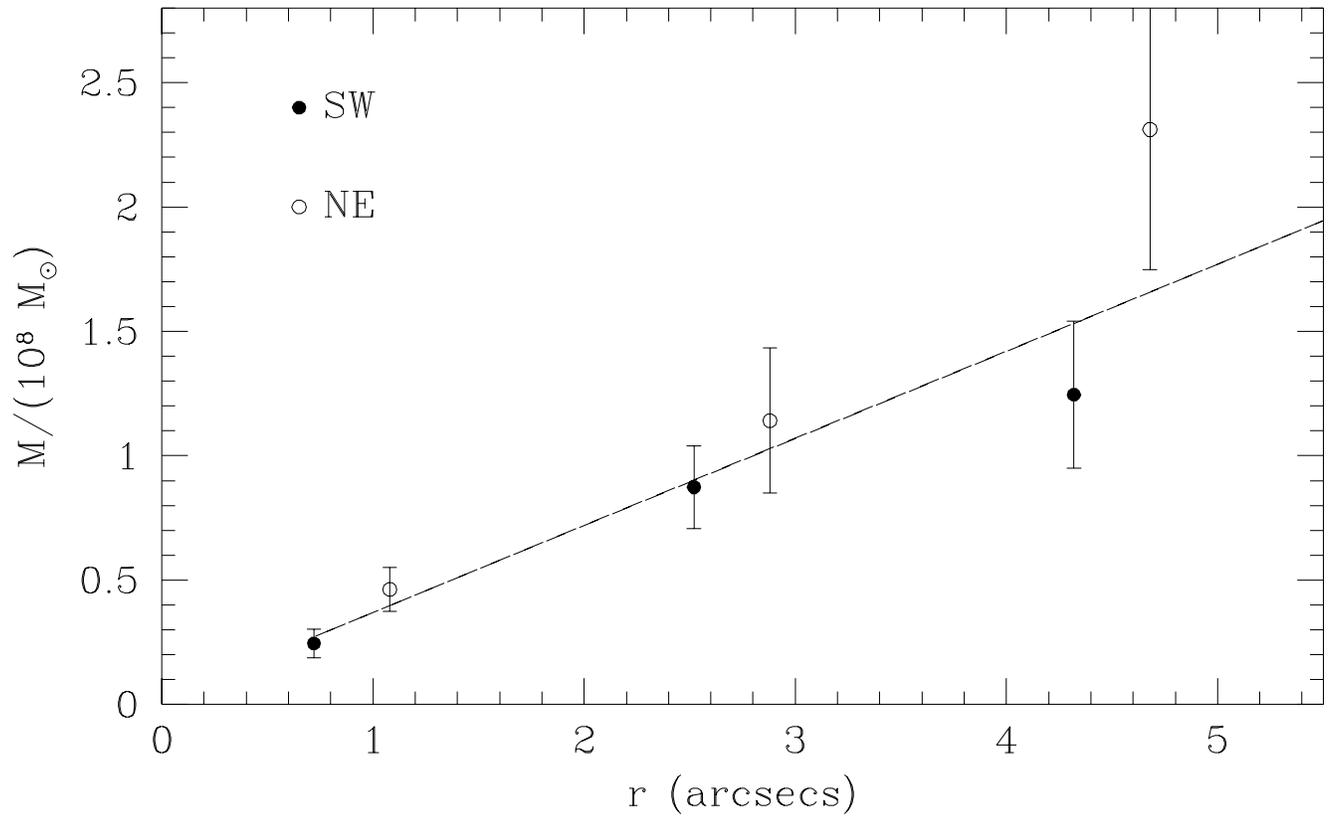
\end{document}